\documentclass[aps,prl,notitlepage,reprint,twocolumn,showpacs,superscriptaddress,nofootinbib,floatfix]{revtex4-1}

\usepackage{amsmath,amssymb,graphicx}
\usepackage{mathrsfs}
\usepackage{bbm}
\usepackage{bbold}
\usepackage{color}
\usepackage{hyperref}

\newcommand{\be}{\begin{equation}}
\newcommand{\ee}{\end{equation}}
\newcommand{\bi}{\begin{itemize}}
\newcommand{\ei}{\end{itemize}}
\newcommand{\bea}{\begin{eqnarray}}
\newcommand{\eea}{\end{eqnarray}}

\newcommand{\vc}[1]{\mbox{\textbf{#1}}}

\newcommand{\pad}[2]{\frac{\partial #1}{\partial #2}}

\definecolor{bc}{rgb}{0, 0.7, 0.0}

\newcommand{\ud}{\mathrm{d}}
\newcommand{\LCm}{{\scriptscriptstyle -}} 
\newcommand{\LCp}{{\scriptscriptstyle +}}

\newcommand{\LCperp}{{\scriptscriptstyle \perp}}

\usepackage[T1]{fontenc} \usepackage[latin1]{inputenc}

\begin{document} 
\title{Exact classical and quantum dynamics in background electromagnetic fields}

\author{Tom Heinzl}
\email{thomas.heinzl@plymouth.ac.uk}
\affiliation{Centre for Mathematical Sciences, University of Plymouth, PL4 8AA, UK}

\author{Anton Ilderton}
\email{anton.ilderton@plymouth.ac.uk}
\affiliation{Centre for Mathematical Sciences, University of Plymouth, PL4 8AA, UK}

\begin{abstract}
Analytic results for (Q)ED processes in external fields are limited to a few special cases, such as plane waves. However, the strong focussing of intense laser fields implies a need to go beyond the plane wave model. By exploiting Poincar\'e symmetry and superintegrability we show how to construct, and solve without approximation, new models of laser-particle interactions. We illustrate the method with a model of a radially polarised (TM) laser beam, for which we exactly determine the classical orbits and quantum wave functions. Including in this way the effects of transverse field structure should improve predictions and analyses for experiments at intense laser facilities.
\end{abstract}
\maketitle
Recent years have seen a flurry of experimental and theoretical activity in investigating the physics of intense laser fields. Such fields interact strongly with any charged or polarisable matter present, which necessitates a non-perturbative approach. This has traditionally~\cite{Reiss:1962,Narozhnyi:1964,Brown:1964zz}, and near universally, been made possible by modelling the laser field as a plane wave~\cite{Volkov:1935}. Within this model one can address longitudinal and temporal pulse shape effects~\cite{Kibble:1965zz,Neville:1971uc}, the detailed analysis of which has been a central issue in the last decade~\cite{Boca:2009zz,Heinzl:2009nd,Heinzl:2010vg,DiPiazza:2011tq,Dinu:2014tsa}. The same model provides the quantum interaction rates for all widely-used particle in cell (PIC) codes~\cite{Gonoskov:2015}. Thus the plane wave model is a crucial input for current strong field physics. However, the model cannot capture intensity gradient effects stemming from the transverse profile of a laser pulse~\cite{Kibble:1966zz, Kibble:1966zza}. Such effects may become sizeable for strong focussing~\cite{Harvey:2016wte}, precisely the method by which current field strengths are reached.

It is thus necessary to go beyond the plane wave model: this is a long-standing, challenging problem. One option is to resort to suitable approximations, such as assuming high centre-of-mass energy in laser particle collisions~\cite{DiPiazza:2013vra,DiPiazza:2016maj}, or one could try to simplify the defining equations, e.g.~using reduction of order~\cite{Raicher:2015ara,Heinzl:2016kzb}. However, it would clearly be advantageous to also have exact solutions, in particular as these are required for a nonperturbative quantum Furry picture~\cite{Furry:1951zz}. We will show here how to generate such solutions.

Recall that charge motion in a plane wave is~\emph{integrable}, i.e.~exactly solvable, due to the existence of three conserved momenta~\cite{Landau:1987}, stemming from the three translational symmetries of a plane wave. Below we will show for the first time that the plane wave has an even larger number of conservation laws, which render charge motion \emph{maximally superintegrable}~\cite{Wojciechowski:1983,Miller:2013}. In general this implies that part, or all, of a classical orbit may be determined algebraically and, crucially, all known such systems are exactly solvable quantum mechanically~\cite{Tempesta:2001}, as is indeed the case for plane waves~\cite{Volkov:1935}. This inspires our approach to constructing new exactly solvable models for laser-matter interactions: we search for superintegrable systems. This is not to say that we search blindly; we will give below a method which automatically generates systems with conserved quantities. We will also illustrate and validate the method by applying it to a toy model of a radially polarized (TM) beam, solving for both the classical particle trajectories and quantum wave functions exactly. We will see that these are as analytically tractability as for plane wave backgrounds, but capture both transverse field effects and pair production. Further examples will be provided in a separate article.

We begin with the classical problem. Let $P_\mu = m\dot{x}_\mu + eA_\mu$ be the canonical particle momenta in a background (e.g.~laser) field $A_\mu$ and let $\xi^\mu$ generate a Poincar\'e transformation, $\xi^\mu \equiv a^\mu + \omega^{\mu\nu}x_\nu$, where $a_\mu$ and $\omega_{\mu\nu}$ ($\omega_{\mu\nu}+ \omega_{\nu\mu}= 0$) parametrize the four translations and six Lorentz transformations of the 10 dimensional Poincar\'e group. The key question to answer is how the background affects the 10 corresponding Noether charges $\xi.P$ which would be conserved for a free particle. Dotting $\xi_\mu$ into the equations of motion $m \ddot{x}^\mu = e F^{\mu\nu} \dot{x}_\nu$ shows that 
\be\label{Q-DOT}
 \frac{\ud}{\ud \tau} \xi.P = e \dot{x}^\mu \mathcal{L}_\xi A_\mu \; ,
\ee
thus the change in time of the free Noether charges is given by the Lie derivative of the background field, $\mathcal{L}_\xi A_\mu = \xi.\partial A_\mu + A_\nu \partial_\mu \xi^\nu$. Suppose then that $A_\mu$ is a \emph{symmetric} gauge field \cite{Jackiw:1995be}, meaning that for a given $\xi$ the Lie derivative vanishes up to gauge transformations:
\be\label{GAUGE}
	\mathcal{L}_\xi A_\mu(x) = \partial_\mu \Lambda(x) \; .
\ee
In this case (\ref{Q-DOT}) can be integrated and implies the existence of a conserved quantity $Q$,
\be
	Q:=\xi.P  - e\Lambda = \text{constant.}
\ee
Thus we have the important result that any Poincaré symmetry of the background yields a conserved quantity in the charge dynamics. With sufficiently many such symmetries we obtain an integrable system. The precise requirements are formulated in the Hamiltonian picture. The covariant Hamiltonian of particle dynamics vanishes, though, due to reparametrisation invariance~\cite{Dirac:1950pj}, so we must gauge fix by choosing a time parameter to describe dynamical evolution~\cite{Dirac:1949cp}. All choices lead to equivalent results, but for the fields we consider the most suitable choice is, as could be expected from the plane wave case~\cite{Neville:1971uc,Bakker:2013cea}, light-front time $x^\LCp=t+z$. Phase space is then six dimensional, spanned by the canonical coordinates $x^\LCm = t - z$, $\vc{x}^\LCperp =(x,y)$ and the corresponding conjugate momenta $P_\LCm = (P_t - P_z)/2$, $\vc{P}_\LCperp = (P_x, P_y)$. The Hamiltonian and equal time ($x^\LCp$) Poisson brackets~are
\be\label{H:LF}
\begin{split}
  H \equiv P_\LCp &= \frac{(\vc{P}_\LCperp - e\vc{A}_\LCperp)^2 + m^2}{4(P_\LCm - e A_\LCm)}
   + e A_\LCp \; , \\
     \{x^\LCm , P_\LCm \} &=  1 \; , \quad \{ x^i, P_j \} = \delta^i_{\;j} \; . 
\end{split}
\ee
The evolution of any quantity $Q$ is now determined by 
\be\label{Q-DOT-LF}
  \frac{\ud Q}{\ud x^\LCp} = \pad{Q}{x^\LCp} -\{Q, H \} \; .
\ee
For the dynamics to be integrable, there must exist three independent charges $Q_i$ which are conserved, $\ud Q_i/\ud x^\LCp=0$, and in involution, $\{Q_i,Q_j\}=0$ for all $i,j$. If there are additional $Q_k$, not necessarily in involution, then the dynamics is superintegrable~\cite{Wojciechowski:1983,Miller:2013}. The key point is that a source of such $Q$ is given by (\ref{GAUGE}).  (There may be further conserved quantities corresponding to non-Poincar\'e symmetries of phase space.)  We illustrate these ideas with the plane wave case. Let $n_\mu$ be a light-like vector, $n^2=0$, and $l^j_\mu$ be the two mutually orthogonal space-like vectors obeying $l^j.n=0$. A plane wave is then described by
\be\label{PW:FA}
	eF_{\mu\nu} = (n l^j - l^j n)_{\mu\nu} f_j'(n.x) \, , \;  eA_\mu = l^j_\mu f_j(n.x) \;,
\ee
where $f$ is an arbitrary profile. The field (\ref{PW:FA}) is symmetric for $\xi_\mu$ equal to $n_\mu$ or $l_\mu^j$; this implies, choosing coordinates such that $n.x=x^\LCp$, that the three particle momenta $P_\LCm$ and $\vc{P}_\LCperp$ are conserved. These are in involution, giving integrability. However it does not previously seem to have been observed that charge motion in plane waves is superintegrable, because $F_{\mu\nu}$ is also symmetric under two null rotations $T_j$, for which $\xi^j_\mu = (n l^j - l^j n)_{\mu\nu} x^\nu$, i.e.
\be
		T_j \equiv \xi^j.P = x^j P^\LCp - x^\LCp P^j \;.
\ee
(Recall $P^\LCp\equiv 2 P_\LCm$ in lightfront~coordinates.) The potential in~(\ref{PW:FA}) is invariant up to a gauge term as in~(\ref{GAUGE}), yielding the two conserved quantities, for $g'_j \equiv f_j$, 
\be\label{T-TILDE}
	Q_j = T_j + g_j(x^\LCp)  = \text{constant}\;,
\ee
as can be verified using (\ref{Q-DOT-LF}).  Together $\{P_\LCm, P_j, Q_j\}$ are five independent, conserved quantities, three in involution, so particle motion is superintegrable. The transverse particle orbit in a plane wave can then be derived \textit{algebraically} just by rearranging (\ref{T-TILDE}), see \cite{HeinzlIlderton:2017}.

\begin{figure}[t!]
\includegraphics[width=0.49\columnwidth]{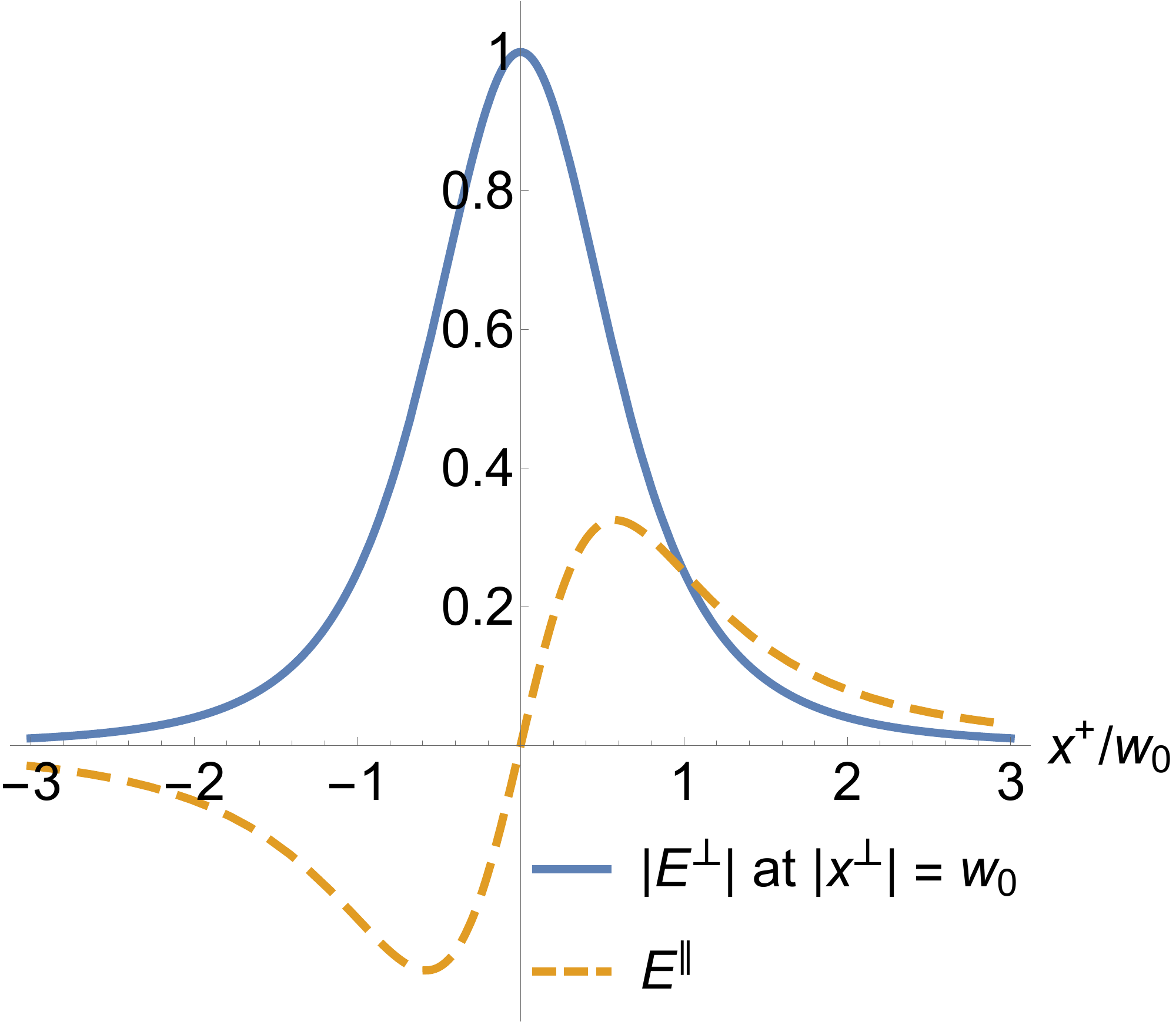}
\includegraphics[width=0.49\columnwidth]{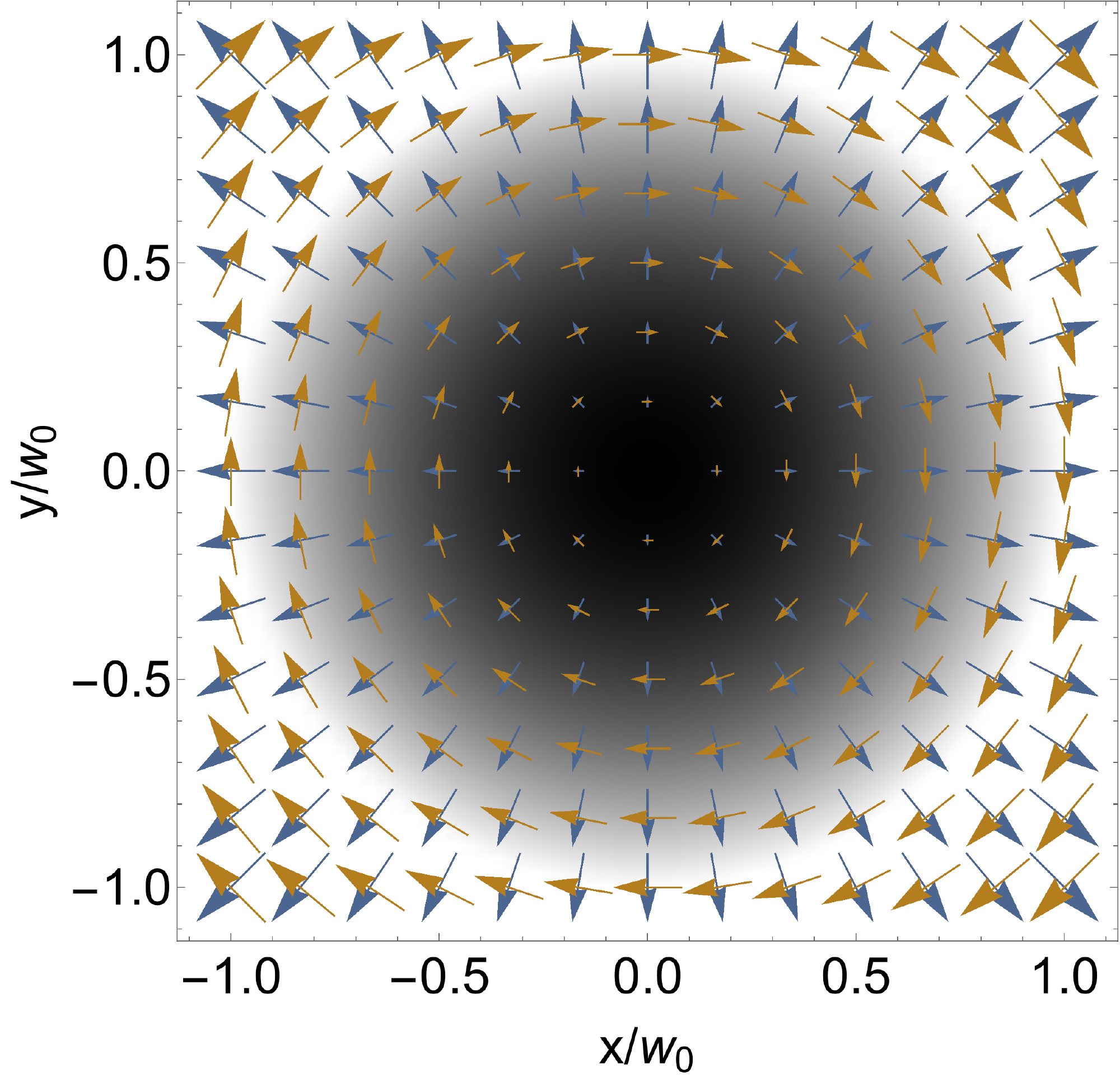} \\\includegraphics[width=0.8\columnwidth]{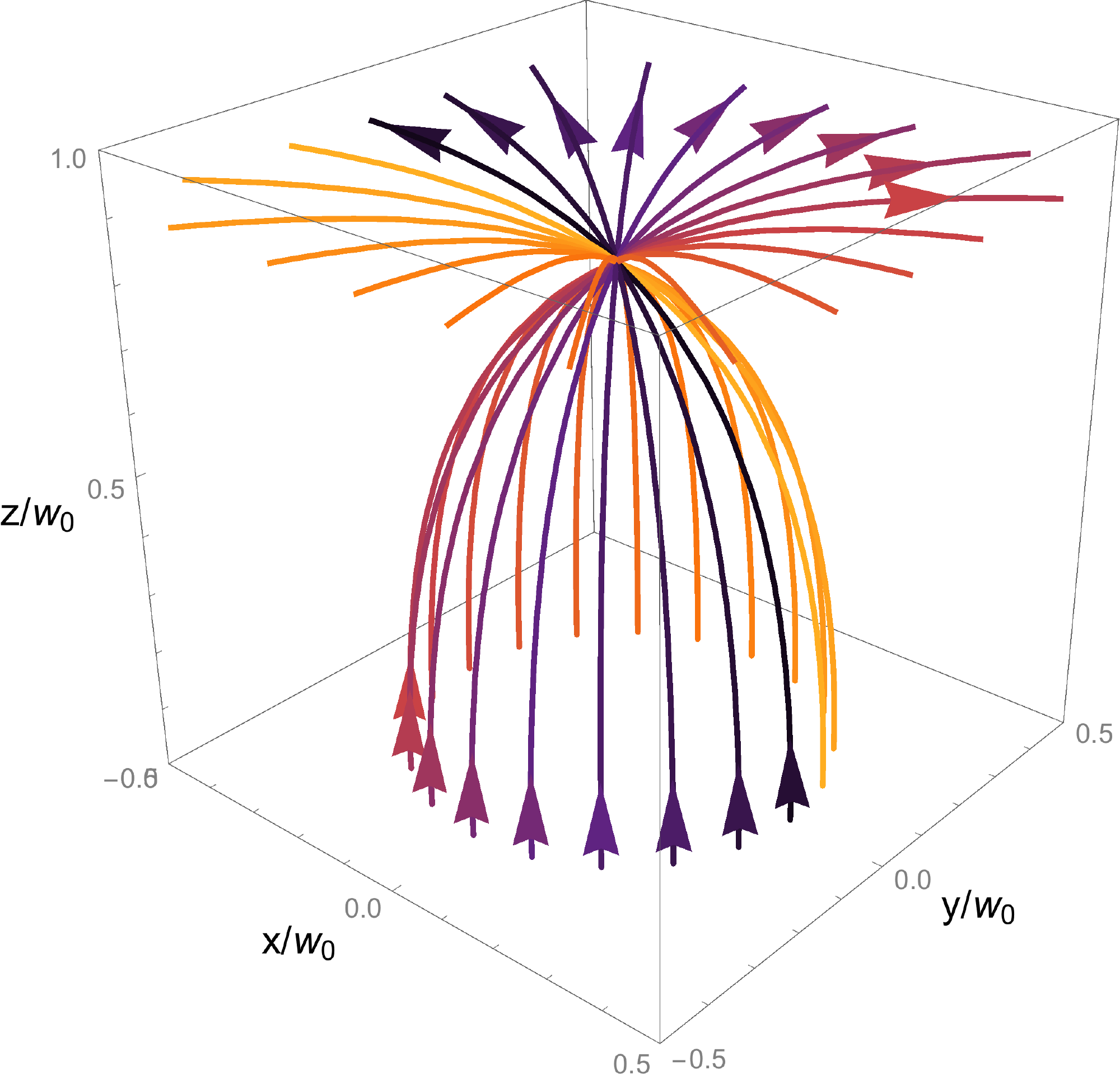}
\caption{\label{FIG:3D} \textit{Upper left:} The transverse and longitudinal electric field profiles (\ref{E-example1}). \textit{Upper right:} density and vector field plot of the transverse fields at fixed $x^\LCp$. \textit{Lower panel:} Electron trajectories at $\gamma=-1/2$. A circular distribution of electrons in the transverse plane, initially at rest when the fields vanish, is drawn toward the symmetry ($z$) axis and focussed to a single point as time evolves. The electrons are expelled from this point as the fields switch off.}
\end{figure}

Before moving on to QED, we present a new example of superintegrable charge motion using a toy model of a radially polarised TM beam (also called a ``doughnut'' or TEM$_{01*}$ mode~\cite{Siegman:1986,Menzel:2007}). Such a field can be generated as the superposition of two linearly polarised Hermite-Gauss TEM$_{01}$ modes~\cite{Kogelnik:1966}, as realised experimentally in~\cite{Oron:2000,Quabis:2005}. Its polarisation structure may be employed to increase focussing~\cite{Dorn:2003}. The TM mode in question has radially (azimuthally) polarised electric (magnetic) fields transverse to the beam propagation direction~\cite{Pohl:1972,Davis:1981,McDonald:2000b}, suggesting rotational symmetry under, say, $L_z$. For a doughnut mode the transverse fields go like~$\sim x^\LCperp \exp -|x^\LCperp|^2$, hence vanish on-axis, rise linearly with $x^\LCperp$ close to the axis, and have a maximum off-axis beyond which they decrease radially. While the magnetic field is strictly transverse, the electric field has a longitudinal component, which has essentially the same temporal dependence as the transverse fields but is suppressed relative to them (in the paraxial approximation by the usual small focussing parameter for a Gaussian beam~\cite{McDonald:2000b}); this connection between transverse and longitudinal profiles suggests retaining invariance under $T_j$. We also retain invariance under $P_\LCm$. A field with the four symmetries $\{P_\LCm,T_j,L_z\}$ is
\be\label{E-B}
\begin{split}
	F_{\mu\nu} &= (x_\mu n_\nu -n_\mu x_\nu) \mathcal{E}(n.x) \;, \\
	{\bf E} = \mathcal{E}(x^\LCp)&\big( x,y,x^\LCp \big) \;, \quad {\bf B} = \mathcal{E}(x^\LCp)\big( y , -x, 0 \big) \;.
\end{split}
\ee
The transverse electric/magnetic field has radial/azimuthal polarisation, as for a TM beam. The profile $\mathcal{E}$ is common to the transverse and longitudinal fields, and the latter is suppressed by the additional factor of $x^\LCp$ near the temporal peak of the field. (If we take~$\mathcal{E}$ to be peaked at the origin, then $x^\LCp \mathcal{E}(x^\LCp) < \mathcal{E}(x^\LCp)$ close to the peak.) Further, the relativistic invariants of (\ref{E-B}) obey ${\bf E}.{\bf B} = 0$ and ${\bf E}^2-{\bf B}^2 > 0$ in agreement with~\cite{McDonald:2000b}. The field (\ref{E-B}) therefore captures some essential features of a TM doughnut mode near the beam axis, including the polarisation structure, the local rise of the transverse fields, and the suppression of the longitudinal field. (See~\cite{BialynickiBirula:2004ev} for a similar approximation based on a vortex beam.) In addition, the fields (\ref{E-B}) obey the homogeneous Maxwell equations. A potential with $\mathcal{L}_\xi A_\mu=0$ for the four Poincar\'e generators above is 
\be
	eA_\mu = \big(\frac{x.x}{2n.x}n_\nu - x_\nu \big) \frac{f(n.x)}{n.x} \;,
\ee
where $f'(x) := x\mathcal{E}(x)$. The background (\ref{E-B}) is not in general source-free, but in this proof-of-principle investigation the key point is the solvability of the \textit{charge dynamics}, to which we now turn. The four quantities $\{P_\LCm,\vc{T}_\LCperp,L_z\}$ are conserved, and the first three are in involution, implying superintegrability of the charge motion. (There is a fifth conserved quantity corresponding to a non-Poincar\'e symmetry; this will be discussed elsewhere.)   The conservation of $\{P_\LCm,\vc{T}_\LCperp\}$ reduces the equations of motion from second to first order; we read off
\bea\label{skog}
	P_\LCm &=& \text{const.} \implies m \dot{x}^\LCp = p^\LCp - eA^\LCp(x^\LCp) \;, \\
	\label{skog2}
	\vc{T}_\LCperp  &=& \text{const.} \implies m \dot{\vc{x}}^\LCperp 
	= p^\LCp \frac{\vc{x}^\LCperp- \vc{x}^\LCperp_0}{x^\LCp} - 
	e \vc{A}^\LCperp,
\eea
where $p^\LCp$ and $\vc{x}^\LCperp_0$ are the conserved quantities. Equation (\ref{skog}) is immediately solved by quadrature:
\be\label{xp-quad}
	 m\int\limits^{x^\LCp} \! \frac{\ud s}{p^\LCp -e A^\LCp(s)} = \tau \;,
\ee
which allows us to trade $\tau$ for $x^\LCp$, as in the plane wave case. Equation (\ref{skog2}) can now be solved to give the transverse orbit as a function of $x^\LCp$, also as for plane waves,
\be\label{perp-general}
	\vc{x}^{\LCperp}(x^\LCp) =  \vc{u}_0^\LCperp x^\LCp - p^\LCp 
	\vc{x}_0^\LCperp x^\LCp \int\limits^{x^\LCp}\frac{\ud s}{s^2(p^\LCp - e A^\LCp(s))} \;,
\ee
where $\vc{u}_0^\LCperp$ can be related to the initial transverse velocity. Having determined $x^\LCp$ and $\vc{x}^\LCperp$, a direct integration yields $x^\LCm$ via the mass-shell condition,
\be
x^\LCm = \int\limits^{x^\LCp}\!\ud x^\LCp\ \frac{\ud x^\LCperp}{\ud x^\LCp} \frac{\ud x^\LCperp}{\ud x^\LCp} + \bigg(\frac{\ud \tau}{\ud x^\LCp}\bigg)^2 \;.
\ee
As for plane waves, the exact solution to the equations of motion is expressed in terms of the potential and integrals over it. These integrals can be performed analytically in certain cases, also as for plane waves. We give an explicit example, such that the electric field has a Lorentz profile,
\be\label{E-example1}
	{\bf E} = \frac{1}{w_0}(x,y,x^\LCp) \frac{E_0}{(1+x^{\LCp2}/w_0^2)^2} \;,
\ee
where $E_0$ has units of the electric field. $w_0$ is a length which, in this near-axis model, can be interpreted as the remnant of the doughnut waist, see the density plot in Fig.~\ref{FIG:3D}. Charge dynamics in this field are controlled by an (inverse) adiabacity parameter $\gamma := eE_0w_0/(2p^\LCp)$. 	
Taking $E_0>0$, we note that $\gamma$ is negative/positive for electrons/positrons. Working with dimensionless $\varphi := x^\LCp/w_0$, the relation~(\ref{xp-quad}) between $x^\LCp$ and $\tau$ becomes, writing $\sqrt{1-\gamma}=:\mu$
\be\label{xplus-gamma-less}
	\frac{p^\LCp}{m} \frac{\tau}{w_0} = \varphi + \frac{\gamma}{\mu} \bigg[\frac{\pi}{2}+ \tan^{-1}\bigg(\frac{\varphi}{\mu} \bigg)\bigg]\;,
\ee
for $-\infty <\gamma < 1$. If $\gamma>1$, the $\varphi$-range accessible by the particle is restricted to either $(-\infty,-\sqrt{\gamma-1})$ or $(\sqrt{\gamma-1},\infty)$ because there are zeros of the denominator in (\ref{xp-quad})--(\ref{perp-general})~\cite{Woodard:2001hi}. The corresponding expressions are given in the appendix. Turning to the transverse coordinates, the explicit solution is, for $0<\gamma<1$, 
\be
\label{xperp-gamma-less}
\vc{x}^\LCperp= \vc{x}_0^\LCperp + \vc{u}_0^\LCperp \varphi + \vc{x}_0^\LCperp\frac{\gamma}{\mu^2} \bigg[1+ \frac{\varphi}{\mu} \bigg(\frac{\pi}{2}+\tan^{-1}\bigg(\frac{\varphi}{\mu} \bigg)\bigg)\bigg] \;.
\ee
The expression for $x^\LCm$ is similar. The electron orbits are illustrated in Fig.~\ref{FIG:3D} and clearly exhibit particle focussing: electrons are drawn toward the beam axis and are focussed to the same spatial position at the same time (while positrons are repelled from the axis, see the appendix). As the fields switch off, the electrons are flung out from the focal point.  These behaviours are due to the transverse field structure: in a plane wave, all electrons would be pushed in the same transverse direction. Thus, even in our simplified field model, we can access physics to which the plane wave approximation is blind. It would be interesting to examine profiles $\mathcal{E}(x^\LCp)$ with multiple field oscillations, which may allow an analytic investigation of particle trapping as well as focussing.

We now extend our method to the Dirac and Klein-Gordon equations in a background $A_\mu$, in order to obtain Furry picture wave functions \cite{Furry:1951zz} for use in QED calculations.  In order to give a unified description we present the scalar and spinor cases together.  It may be verified directly from the equations of motion that the field theory version of (\ref{Q-DOT}) is
\be\begin{split}\label{CURRENT1}
	\partial_\mu\big[ (D^\mu\phi)^\dagger \mathcal{L}_\xi\phi + \text{c.c.}\big] &= ej^\mu \mathcal{L}_\xi A_\mu \,, \\
	\partial_\mu\big[ i\bar\psi \gamma^\mu \mathcal{L}_\xi \psi \big] &= ej^\mu \mathcal{L}_\xi A_\mu \,.
\end{split}
\ee
Here the scalar and spinor Lie derivatives corresponding to Poincar\'e transformations are
\be\begin{split}
	\mathcal{L}_\xi \phi &= \xi.\partial \phi \;, \quad \mathcal{L}_\xi \psi = \xi.\partial \psi +\frac{1}{2}\partial_\mu\xi_\nu \big[\gamma^\mu, \gamma^\nu\big]\psi\; .
\end{split}
\ee
The U(1) currents in (\ref{CURRENT1}), $j^\mu=\bar\psi \gamma^\mu \psi$ for spinors and $j^\mu = i\phi^\dagger D^\mu\phi +\text{c.c.}$ for scalars, are conserved, $\partial_\mu j^\mu = 0$. If the potential is symmetric, $\mathcal{L}_\xi A_\mu = \partial_\mu \Lambda$, then from (\ref{CURRENT1}) we also have conserved Poincar\'e currents,
\be\begin{split}\label{CURRENT2}
	J^\mu_\xi &= (D^\mu\phi)^\dagger [\mathcal{L}_\xi \phi +ie\Lambda\phi]+ \text{c.c.} \;, \\
	J^\mu_\xi &= i\bar\psi \gamma^\mu\big[\mathcal{L}_\xi \psi +i e\Lambda \psi\big] \;. 
\end{split}
\ee
We now employ this symmetry to solve for $\phi$ or $\psi$. To this end we note that a sufficient condition guaranteeing conservation of (\ref{CURRENT2}) is
\be\begin{split}\label{LOCAL}
	(i \mathcal{L}_\xi -e\Lambda)\phi &= \lambda \phi \;, \quad (i\mathcal{L}_\xi -e \Lambda)\psi = \lambda\psi\;,
\end{split}
\ee
with real $\lambda$, for then $\partial_\mu j^\mu_\xi \propto \partial_\mu j^\mu = 0$. We then impose (\ref{LOCAL}) for as many ``$i\mathcal{L}_\xi -e\Lambda$'' as possible -- these must clearly be mutually commuting, which is the analogue of the classical involution condition. The idea is that, as all known maximally superintegrable classical systems are also solvable quantum mechanically~\cite{Tempesta:2001,Miller:2013}, this procedure will identify enough of the field structure to make the Dirac/ Klein-Gordon equation soluble. This is exactly what happens for plane waves -- imposing (\ref{LOCAL}) for the three commuting derivatives $\{\partial_\LCm,\partial_\LCperp\}$ reduces e.g.\ the Klein-Gordon PDE to a soluble first order ODE in $x^\LCp$, yielding the Volkov solutions. Let us now add to these results, turning to the TM beam model~(\ref{E-B}).

For simplicity we look for a scalar $\phi$ which is a simultaneous eigenvector of the three commuting operators $\{P_\LCm,T_j\}$. As for a plane wave, this reduces the Klein Gordon equation to an ODE in $x^\LCp$, which is solved by 
\be\label{phi1}
	\phi(x) = \frac{\exp\big(-i \mathcal{S}\big)}{x^\LCp\sqrt{p^\LCp -e A^\LCp(x^\LCp)}} \;,
\ee
where $\mathcal S$ is the classical Hamilton-Jacobi action obeying $(\partial_\mu {\mathcal S} - e A_\mu)^2 = m^2$,
\be\nonumber
	\mathcal{S} := \frac{p^\LCp}{2} \frac{x.x}{x^\LCp}
	+ \frac{p^\LCp |x_0^\LCperp|^2}{x^\LCp} 
	+ \int\limits^{x^\LCp}\!\ud s \frac{p^{\LCp 2}|x_0^\LCperp|^2 
	+ m^2 s^2}{2s^2(p^\LCp-e A^\LCp(s))} \;.
\ee
It is worth highlighting the similarities and differences between (\ref{phi1}) and the Volkov solutions~\cite{Volkov:1935}. The classical action appears in the exponent in both cases, and contains three degrees of freedom. Re-exponentiating the denominator in (\ref{phi1}), we see that the exponent differs from (minus $i$ times) the classical action by an imaginary part, representing quantum corrections. The Volkov solution, on the other hand, is equal to the exponent of the classical action, i.e.~receives no quantum corrections. 

A key difference is the non-trivial dependence of (\ref{phi1}) on the transverse coordinates, due to the transverse field structure. Here it is revealing to compare the Fourier transforms of (\ref{phi1}) and the Volkov solution, 
\begin{widetext}
\begin{align}\label{Fourier-comp}
	\phi(x) = \displaystyle\int\!{\bar \ud} p \, \frac{e^{-ip.x}}{\sqrt{p^\LCp - e A^\LCp(x^\LCp)}}\int\!\ud^2 x_0^\LCperp \Phi(x_0^\LCperp,p^\LCp) \, \exp &\bigg[ i x_0^\LCperp p_\LCperp -\frac{i}{2p^\LCp}\int\limits_{-\infty}^{x^\LCp}\!\ud s \, \frac{eA^\LCp(s)}{p^\LCp - eA^\LCp(s)} \bigg(m^2+\frac{p^{\LCp2} x_0^\LCperp x_0^\LCperp}{s^2}\bigg)\bigg] \;, \\
	 \displaystyle
	\label{Volkov}\phi_\text{Volkov}(x) = \displaystyle\int\!{\bar \ud} p \, \frac{e^{-ip.x}}{\sqrt{p^\LCp}} \Phi(p_\LCperp,p^\LCp) \, \exp &\bigg[ -\frac{i}{2p^\LCp}\int\limits_{-\infty}^{x^\LCp}\!\ud s \ 2 e A^\LCperp(s)p_\LCperp - e^2 A^\LCperp(s) A_\LCperp(s)\bigg] \;,
\end{align}
\end{widetext}
in which $\Phi$ is an arbitrary function in both cases,  $p_\mu$ is on-shell and ${\bar \ud} p$ is the on-shell measure. It is now simple to read off the past asymptotic behaviour of the fields: they become free, with $\Phi$ being the initial wave packet.

Another difference is the presence of $p^\LCp -e A^\LCp$ in denominators, rather than just $p^\LCp$ as appears in the Volkov solution (\ref{Volkov}). This is due to the longitudinal field. The denominator can vanish which, as for a purely longitudinal electric field~\cite{Tomaras:2001vs,Ilderton:2014mla}, reflects the fact that the field can spontaneously produce pairs, as $E^2 - B^2 > 0$, recall the discussion below (\ref{E-B}). Thus our exact results describe classical dynamics, quantum corrections, and pair production effects.

It will be very interesting to investigate the role played by the above symmetries in pair production, nonlinear Compton scattering, and other quantum processes. On this note, let us address the feasibility of calculating with the solutions (\ref{Fourier-comp}). Recall that using Volkov solutions one can always perform the $x^\LCm$ and $x^\LCperp$ integrals at each QED vertex exactly, yielding three delta functions. In the TM case, the $x^\LCm$ integrals will still yield delta functions, and since $\phi$ is Gaussian in $x^\LCperp$ we will still be able to perform all transverse integrals exactly (to yield Gaussians). Hence scattering calculations in the TM background will allow as much analytic progress as the plane wave case. This will be pursued in a future paper, but as a first example the appendix contains a calculation of the influence of the TM field on wavepacket spreading.

In conclusion, we have shown how to construct new, exactly solvable models of classical and quantum charge dynamics in background fields, based on superintegrability. This finally provides a way to go beyond the plane wave model in laser-matter interactions, without approximation. We have demonstrated the feasibility of the method using a first example in which the effects of transverse field structure, to which the plane wave model is blind, can be analytically investigated. This also adds to the mathematical literature of known superintegrable systems. We hope that our exact analytic method will lead to new and improved predictions and analyses for upcoming high intensity laser experiments. Of course our method is not limited to laser fields; another target for investigation is the effect of spatial inhomogeneities in magnetic fields~\cite{Gies:2013yxa}.

\acknowledgments

\textit{We thank Ben King and David McMullan for useful discussions. This project has received funding from the European Union's Horizon 2020 research and innovation programme under the Marie Sk\l odowska-Curie grant No.~701676.}

\appendix

\clearpage
\section{Appendix: positron motion}
\begin{figure}[t!!]
\includegraphics[width=\columnwidth]{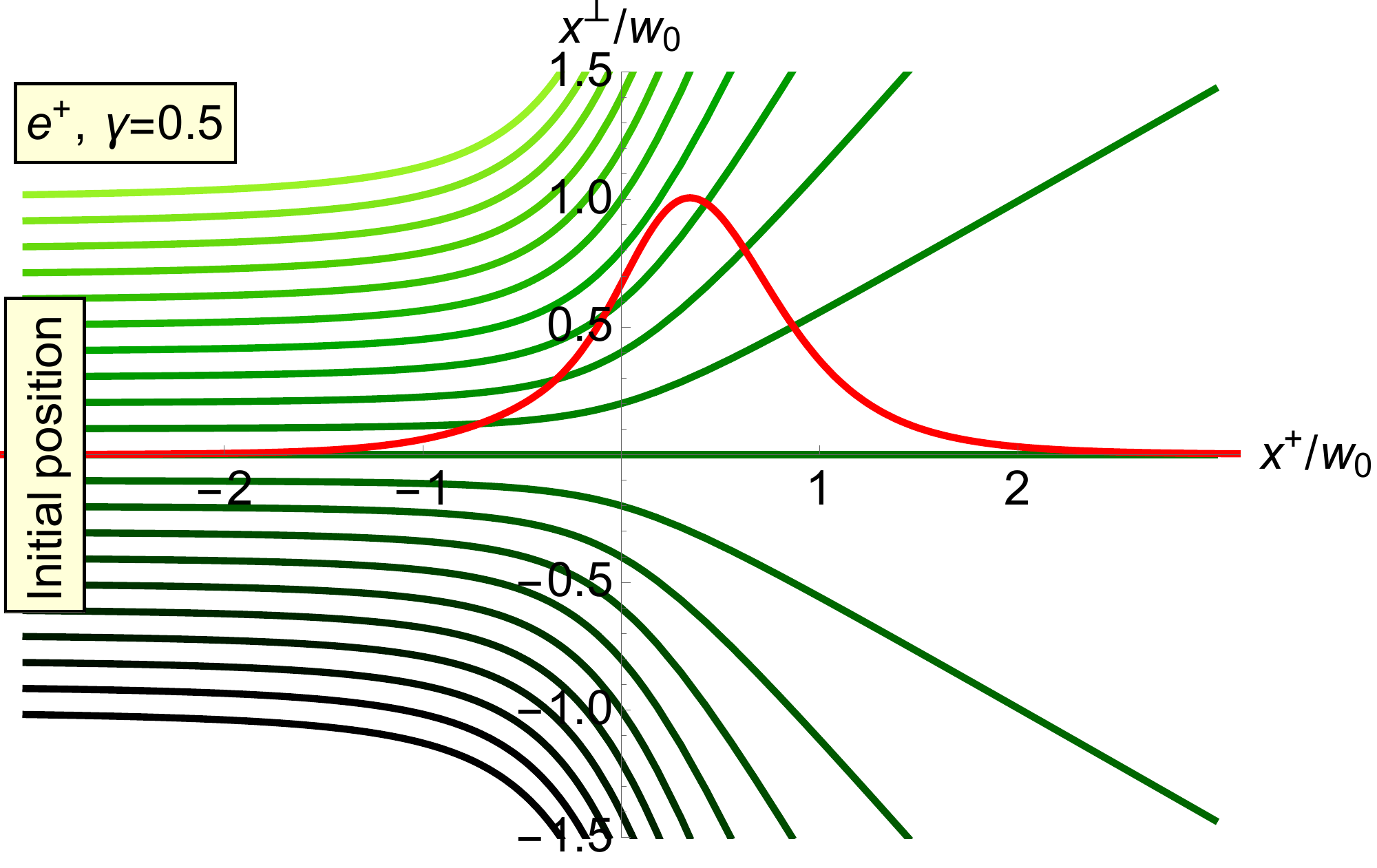}
\caption{\label{FIG:PERP1} 
Transverse motion of positrons at $\gamma = 1/2$. The red curve shows a typical example of the local energy density $(E^2+B^2)/2$ on the positron's trajectory (not to scale), illustrating that despite the transverse rise of the field, the particles are asymptotically free. This can be verified from the asymptotic expansions of (\ref{xplus-gamma-less})--(\ref{xperp-gamma-less}).}
\end{figure}
For completeness we give here the extension of the classical motion results (\ref{xplus-gamma-less})-(\ref{xperp-gamma-less}) to positrons. For $0<\gamma<1$ the explicit expressions are as in the text, and the orbits are illustrated in Fig.~\ref{FIG:PERP1} above. The explicit expressions differ though for $\gamma>1$. For $x^\LCp$ we have, writing $\nu:=\sqrt{\gamma-1}$
\be\label{xplus-gamma-larger}
	\frac{p^\LCp}{m} \frac{\tau}{w_0} = \varphi - \frac{\gamma}{\nu} \coth^{-1}\bigg[\frac{\varphi}{\nu}\bigg] \;,
\ee
and for the transverse coordinates
\be
\label{xperp-gamma-larger}
	\vc{x}^\LCperp = \vc{x}_0^\LCperp + \vc{u}^\LCperp \varphi - 
	\vc{x}_0^\LCperp \frac{\gamma}{\nu^2}  \bigg(1- \frac{\varphi}{\nu}
	 \coth^{-1}\bigg[\frac{\varphi}{\nu} \bigg]\bigg) \;.
\ee

\section{Appendix: Wave packet spreading}
\begin{figure}[t!!!]
\includegraphics[width=\columnwidth]{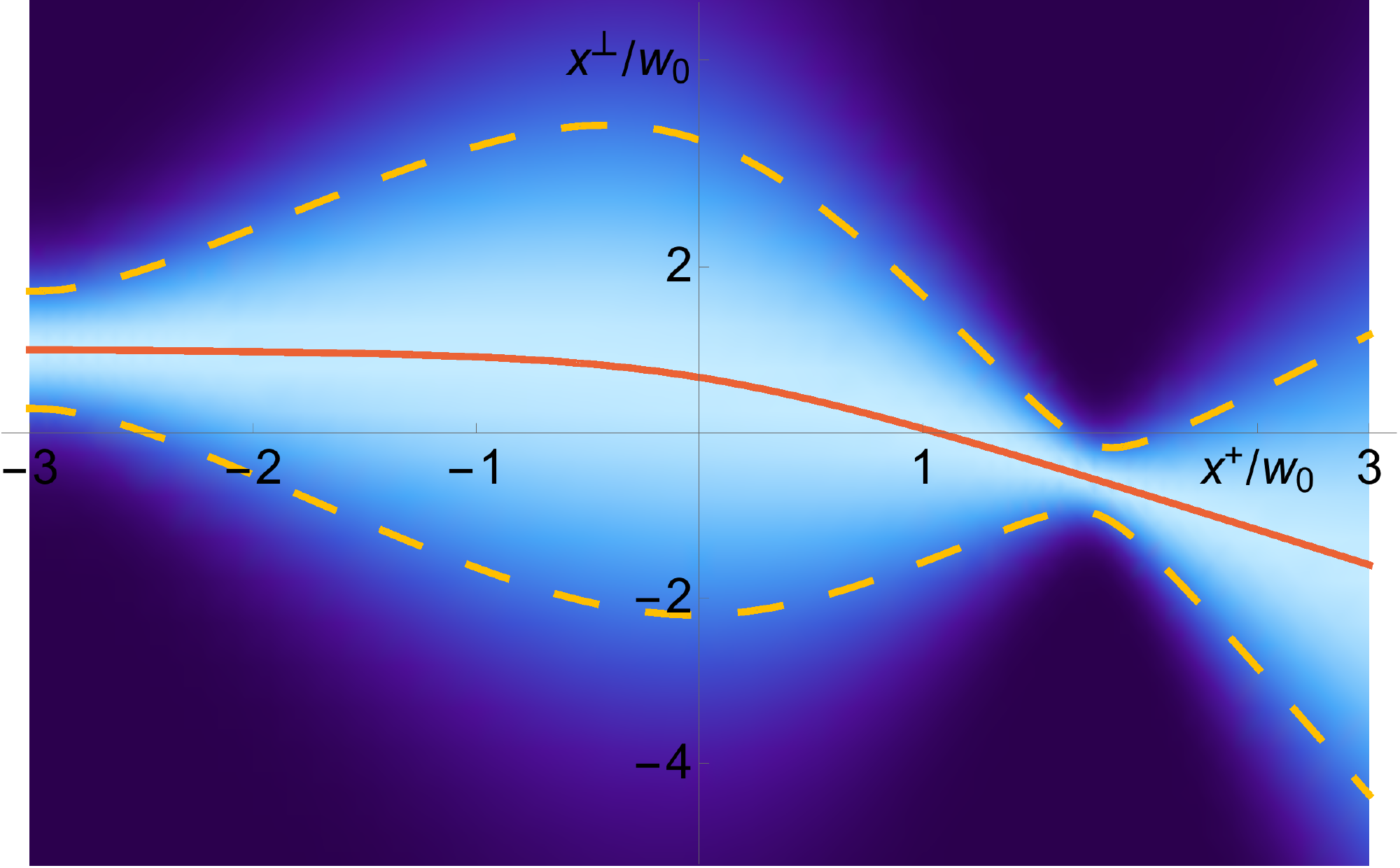}
\caption{\label{FIG:PACKET}  Dynamics of a quantum wave packet of Klein-Gordon solutions. Colour (dark to light blue) shows the density of the Gaussian wave packet, which is centred on the classical trajectory (solid line, initial conditions $p^\LCp=m$, $x^\LCperp_0=w_0$). The dotted line shows $\pm$ one standard deviation $\sigma$ around the classical trajectory ($\sigma_0=w_0/\sqrt{2}$). The wave packet spreads in coordinate space when the field is weak, but can be narrowed by the fields of the TM beam.}
\end{figure}

As a first illustration of a quantum calculation, we show how the TM beam influences the spread of a wave packet of Klein-Gordon solutions. This `first quantised' calculation is only meant to demonstrate the feasibility of using the wave functions (\ref{Fourier-comp}); we concentrate only on the transverse coordinates and the exponential part of the wave function, neglecting pair production.

We choose $\Phi$ in (\ref{Fourier-comp}) such that the real part of the scalar $\phi$ is, at initial time $x^\LCp_0$, Gaussian in the transverse co-ordinates, centred on position $x^\LCperp_0$ and with width (standard deviation) $\sigma_0$. For $x^\LCp > x^\LCp_0$, the real part of the wave packet has the form,
\be
	\phi(x) \sim \exp \bigg[ -\frac{(\vc{x}^\LCperp - \vc{x}^\LCperp_\text{cl})^2}{2\sigma^2}\bigg] \;,
\ee
which is centred on the classical orbit $\vc{x}^\LCperp_\text{cl}$ as in~(\ref{perp-general}), and where the variance of the spreading wavepacket is
\be
	\sigma^2(x^\LCp) = \sigma_0^2(1-x^\LCp \mathcal{I}) + \frac{(x^\LCp-x^\LCp_0+x^\LCp x^\LCp_0 \mathcal{I})^2}{p^{\LCp2}\sigma_0^2} \;,
\ee
with
\be
	\mathcal{I}\equiv \int\limits_{x^\LCp_0}^{x^\LCp}\!\ud s \, \frac{1}{s^2}\frac{eA^\LCp(s)}{p^\LCp - eA^\LCp(s)} \;.
\ee
The width obeys $\sigma(x^\LCp_0)=\sigma_0$ and for $A_\mu\to 0$ recovers the usual free-field spreading of the wave packet,
\be
	\sigma^2(x^\LCp)\bigg|_{A=0} = \sigma_0^2 + \frac{(x^\LCp-x^\LCp_0)^2}{p^{\LCp2}\sigma_0^2} \;.
\ee
Hence when the field is weak the wave packet spreads as in quantum mechanics. The wave packet can though be focussed by the TM field, as shown in Fig.~\ref{FIG:PACKET}. Note that the derivation of these results requires performing Gaussian integrals only. 
\newpage
\end{document}